\documentclass[preprint2]{aastex}

\shorttitle{\indent \def Sunspot oscillations observed with IRIS} \shortauthors{Tian et al.}

\begin{document}

\title{High-resolution observations of the shock wave behavior for sunspot oscillations with the Interface Region Imaging Spectrograph}

\author{H. Tian\altaffilmark{1}, E. DeLuca\altaffilmark{1}, K. K. Reeves\altaffilmark{1}, S. McKillop\altaffilmark{1}, B. De Pontieu\altaffilmark{2}, J. Mart\'{\i}nez-Sykora\altaffilmark{2,3}, M. Carlsson\altaffilmark{4}, V. Hansteen\altaffilmark{4}, L. Kleint\altaffilmark{2,3,5}, M. Cheung\altaffilmark{2}, L. Golub\altaffilmark{1}, S. Saar\altaffilmark{1}, P. Testa\altaffilmark{1}, M. Weber\altaffilmark{1}, J. Lemen\altaffilmark{2}, A. Title\altaffilmark{2}, P. Boerner\altaffilmark{2}, N. Hurlburt\altaffilmark{2}, T. D. Tarbell\altaffilmark{2}, J. P. Wuelser\altaffilmark{2}, C. Kankelborg\altaffilmark{6}, S. Jaeggli\altaffilmark{6}, S. W. McIntosh\altaffilmark{7}}
\altaffiltext{1}{Harvard-Smithsonian Center for Astrophysics, 60 Garden Street, Cambridge, MA 02138; hui.tian@cfa.harvard.edu}
\altaffiltext{2}{Lockheed Martin Solar and Astrophysics Laboratory, 3251 Hanover St., Org. ADBS, Bldg. 252, Palo Alto, CA  94304}
\altaffiltext{3}{Bay Area Environmental Research Institute, 596 1st St West, Sonoma, CA 95476}
\altaffiltext{4}{Institute of Theoretical Astrophysics, University of Oslo, P.O. Box 1029, Blindern, NO-0315 Oslo, Norway}
\altaffiltext{5}{NASA Ames Research Center, Moffett Field, CA 94305}
\altaffiltext{6}{Department of Physics, Montana State University, Bozeman, P.O. Box 173840, Bozeman, MT 59717}
\altaffiltext{7}{High Altitude Observatory, National Center for Atmospheric Research, P.O. Box 3000, Boulder, CO 80307}

\begin{abstract}
We present first results of sunspot oscillations from observations by the Interface Region Imaging Spectrograph (IRIS). The strongly nonlinear oscillation is identified in both the slit-jaw images and the spectra of several emission lines formed in the transition region and chromosphere. We first apply a single Gaussian fit to the profiles of the Mg~{\sc{ii}}~2796.35\AA{}, C~{\sc{ii}}~1335.71\AA{} and Si~{\sc{iv}}~1393.76\AA{} lines in the sunspot. The intensity change is $\sim$30\%. The Doppler shift oscillation reveals a sawtooth pattern with an amplitude of $\sim$10~km~s$^{-1}$ in Si~{\sc{iv}}. In the umbra the Si~{\sc{iv}} oscillation lags those of  C~{\sc{ii}} and Mg~{\sc{ii}} by $\sim$3 and $\sim$12 seconds, respectively. The line width suddenly increases as the Doppler shift changes from redshift to blueshift. However, we demonstrate that this increase is caused by the superposition of two emission components. We then perform detailed analysis of the line profiles at a few selected locations on the slit. The temporal evolution of the line core is dominated by the following behavior: a rapid excursion to the blue side, accompanied by an intensity increase, followed by a linear decrease of the velocity to the red side. The maximum intensity slightly lags the maximum blue shift in Si~{\sc{iv}}, whereas the intensity enhancement slightly precedes the maximum blue shift in Mg~{\sc{ii}}. We find a positive correlation between the maximum velocity and deceleration, a result that is consistent with numerical simulations of upward propagating magneto-acoustic shock waves.
\end{abstract}


\section{Introduction}
Solar magneto-hydrodynamic waves have been intensively studied both observationally and theoretically in the past decades since they are believed to play a crucial role in chromospheric and coronal heating. Moreover, the generation and propagation of these waves can provide valuable information on the thermal and magnetic structures of the solar atmosphere.

Sunspot oscillations, discovered by \cite{Beckers1969}, are one of the most spectacular wave phenomena in the solar atmosphere. Sunspot oscillations often have a dominant period of five minutes at the photospheric level \citep[e.g.,][]{Beckers1972}. These 5-minute oscillations are the response of the sunspot to forcing by the 5-minute p-mode oscillations in the surrounding atmosphere \citep[e.g.,]{Thomas1985}. Three-minute oscillations have been frequently reported in both the photosphere and chromosphere of the sunspot umbrae. Compared to the photospheric lines, they are more easily observed in the brightness (also called umbral flashes) and velocity of the chromospheric Ca~{\sc{ii}} H\&K lines, as well as in velocities derived from the He~{\sc{ii}} 10830\AA{} triplet. Early theoretical efforts suggest that they are a resonant mode of the sunspot itself \citep[see a review in][]{Thomas1985}. However, the detection of 3-minute oscillations in the transition region (TR) and corona above sunspots supports the interpretation of propagating waves \citep[e.g.,][]{Brynildsen1999a,Brynildsen1999b,Brynildsen2002,Brynildsen2004,Maltby1999,OShea2002,DeMoortel2002}. Running penumbral waves (RPWs) are concentric acoustic waves propagating outward with a speed of 10-25 km~s$^{-1}$ from the umbra-penumbra boundary \citep[e.g.,][]{Giovanelli1972,Zirin1972}. Their periods are usually in the range of 200-300 seconds. The RPWs are suggested to be trans-sunspot waves of purely chromospheric origin \citep[e.g.,][]{Tziotziou2006} or magneto-acoustic waves propagated along expanding field lines from the photosphere \citep[e.g.,][]{Bloomfield2007,Jess2013}.  For reviews of the observations and theories of sunspot oscillations, we refer to \cite{Lites1992}, \cite{Bogdan2000}, and \cite{Bogdan2006}.

As pointed out by \cite{Bogdan2000}, our most profound ignorance centers on the nonlinear aspects of the sunspot oscillations. Although signatures of nonlinearity and upward propagating shock waves have been frequently reported in chromospheric lines \citep[e.g.,][]{Bard1997,Rouppe2003,Centeno2006,Felipe2010,delaCruz2013}, direct evidence of the shock wave nature for sunspot oscillations in the TR is very rare. For instance, \cite{Brynildsen1999a} did not find any clear signs of shocks for the sunspot oscillations in their moderate-resolution observations. \cite{OShea2002} mentioned that it was not possible to determine whether the oscillations they observed were linear or non-linear waves due to the poor resolution of the instrument. Nonlinear sunspot oscillations were identified and explained as nonlinear acoustic waves without shocks by \cite{Brynildsen1999b} and \cite{Brynildsen2004}, although Brynildsen et al. (1999b) suggested the possible presence of shocks.

The recently launched Interface Region Imaging Spectrograph \citep[IRIS,][]{DePontieu2014} mission is now providing high-cadence, high-resolution, and continuous observations of the solar TR and chromosphere. Here we report the first result of sunspot oscillations observed with IRIS. The new IRIS observations provide direct and adequate evidences of the shock wave nature for sunspot oscillations in the TR and chromosphere. These evidences include a sharp change of the velocity and a clear correlation between the maximum velocity and deceleration.

\section{Data analysis}

\begin{figure*}
\centering {\includegraphics[width=\textwidth]{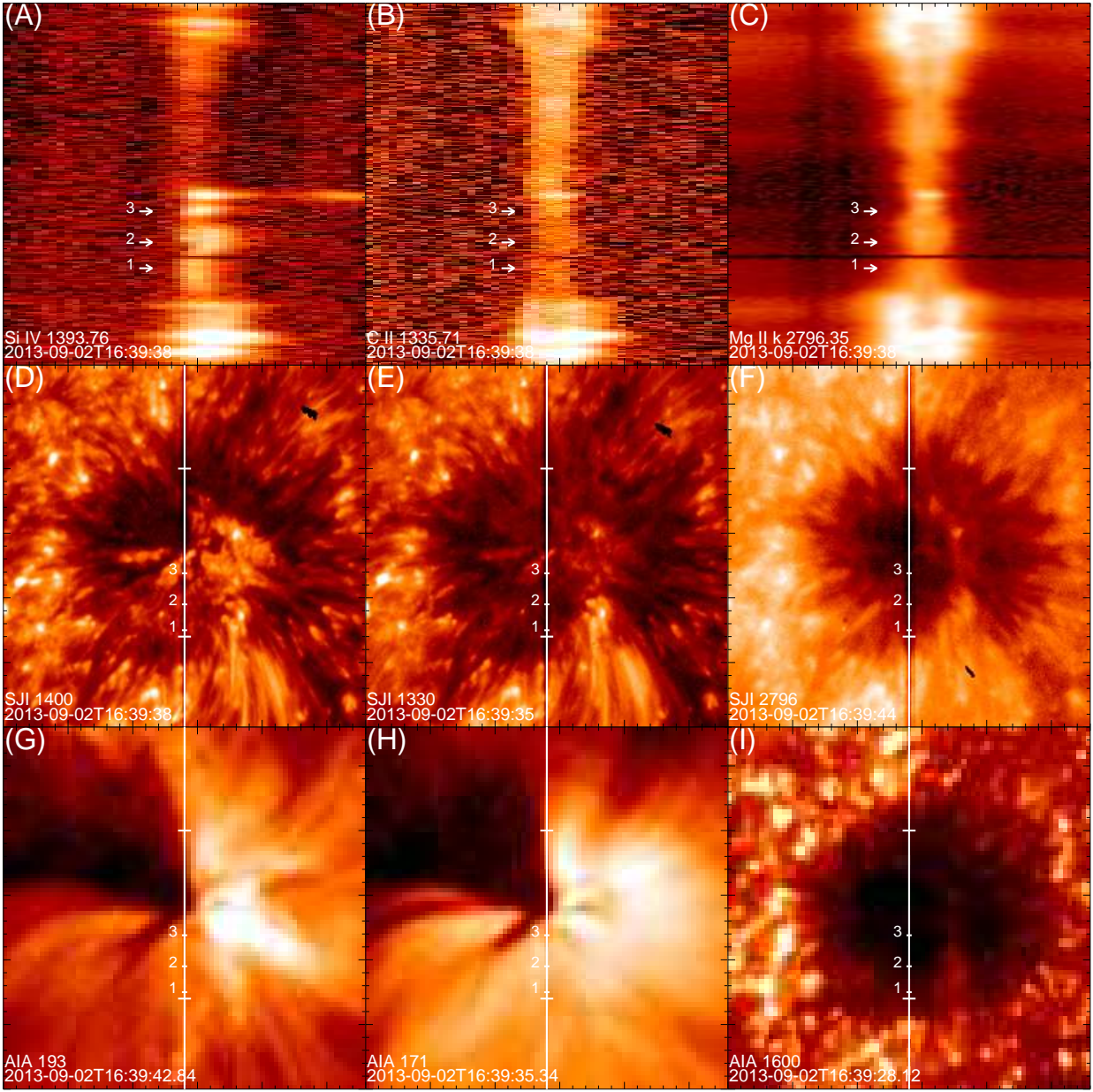}} \caption{ Images and spectra around 16:39 UT on 2013 September 2. (A)-(C) Si~{\sc{iv}}~1393.76\AA{}, C~{\sc{ii}}~1335.71\AA{} and Mg~{\sc{ii}}~2796.35\AA{} spectra along the slit. The length along the slit (vertical dimension) is 47$^{\prime\prime}$. The wavelength range (horizontal dimension) is about 1.2\AA{} for the FUV lines and 2.4\AA{} for Mg~{\sc{ii}}. (D)-(F): IRIS SJI images in different filters. (G)-(I): AIA 193\AA{}, 171\AA{} \& 1600\AA{} images. The field of view of the SJI and AIA images has a size of 47$^{\prime\prime}$$\times$47$^{\prime\prime}$. In (D)-(I) the vertical white line marks the location of the IRIS slit and the two long horizontal lines indicate the spatial range shown in Figure~\ref{fig.2}. The three short horizontal lines in (D)-(I) mark the slit locations (labeled 1, 2 and 3) where we perform detailed analysis of the line profiles. The corresponding spectral features are pointed by the three arrows in (A)-(C). Two online movies are associated with this figure. }
\label{fig.1}
\end{figure*}

We analyze the sit-and-stare observation made from 16:39 to 17:59 on 2013 September 2. The slit was centered at (99$^{\prime\prime}$, 58$^{\prime\prime}$). The spatial pixel size is 0.167$^{\prime\prime}$. The cadence of the spectral observation in both the near ultraviolet (NUV, 2783-2834\AA{}) and far ultraviolet (FUV, 1332-1358\AA{} \& 1390-1406\AA{}) wavelength bands was 3 seconds. Exposure times were 2 seconds. Slit-jaw images (SJI) in the filters of 2796\AA{}, 1400\AA{} and 1330\AA{} were taken at a cadence of 12 seconds. The calibrated level 2 data was used in our study. Dark current subtraction, flat field correction, and geometrical correction have been taken into account in the level 2 data \citep{DePontieu2014}.

The orbital variation of the line positions includes mainly two components resulting respectively from the temperature change of the detector and spacecraft-Sun distance change over the course of an orbit. The latter has the same effect on both the NUV and FUV spectra and can be monitored using the house keeping data. The former has been found to be negatively correlated between FUV and NUV lines and can be evaluated using a strong NUV line Ni~{\sc{i}}~2799.474\AA{}. 

We mainly use three strong lines for this study: Si~{\sc{iv}}~1393.76\AA{}, C~{\sc{ii}}~1335.71\AA{} and Mg~{\sc{ii}}~K~2796.35\AA{} formed in the middle TR (formation temperature 10$^{4.9}$K), lower TR (10$^{4.4}$K) and chromosphere (10$^{4.0}$K), respectively. Figure~\ref{fig.1} shows the spectral images as well as the context SJI images taken around 16:39 UT. The 2796\AA{} filter samples emission mainly from the Mg~{\sc{ii}}~K~2796.35\AA{} line, while emission in the 1330\AA{} and 1400\AA{} filters comes from both TR lines and UV continuum. The coalignment between different optical channels (three wavelength bands and different SJI filters) was achieved by checking the position of the horizontal fiducial line and the locations of some dynamic events. Figure~\ref{fig.1}(G)-(I) show the images of the 193\AA{}, 171\AA{} and 1600\AA{} passbands taken by the Atmospheric Imaging Assembly \citep[AIA,][]{Lemen2012} onboard the Solar Dynamics Observatory \citep[SDO,][]{Pesnell2012}. Cross-correlation between the AIA 1600\AA{}~image and the IRIS SJI 2796\AA{}~image were used for the coalignment.  

From the online movies we can clearly see the spectacular sunspot oscillation in both the three emission lines and the SJI images. In SJI 2796\AA{} we see clear emission from the umbral flashes. In SJI 1330\AA{} and 1400\AA{} the oscillations seem to be dominated by running waves. The signal-to-noise ratio (S/N) of the Si~{\sc{iv}}~line is low in the north part of the sunspot. Nevertheless, very bright oscillating Si~{\sc{iv}}~features can be clearly seen in the south part of the sunspot. Comparison with the AIA images suggests that these bright TR structures correspond to the footpoints of many coronal loops. The oscillating Si~{\sc{iv}}~emission features seem to be associated with the bottom part of the sunspot plumes which are basically the legs of coronal loops originating from the umbrae \citep[e.g.,][]{Foukal1974,Tian2009}. 

Another interesting phenomenon is the secondary emission peak of Si~{\sc{iv}}~in the middle of Figure~\ref{fig.1}(A). This feature represents a persistent TR downflow of $\sim$98 km~s$^{-1}$ in the umbra, likely supersonic. Such downflows were previously reported by \cite{Brekke1993} and \cite{Brynildsen2001}. From the online movies we do not see apparent oscillation of this downflow. 

\begin{figure*}
\centering 
\begin{minipage}[t]{0.6\textheight}
{\includegraphics[width=\textwidth]{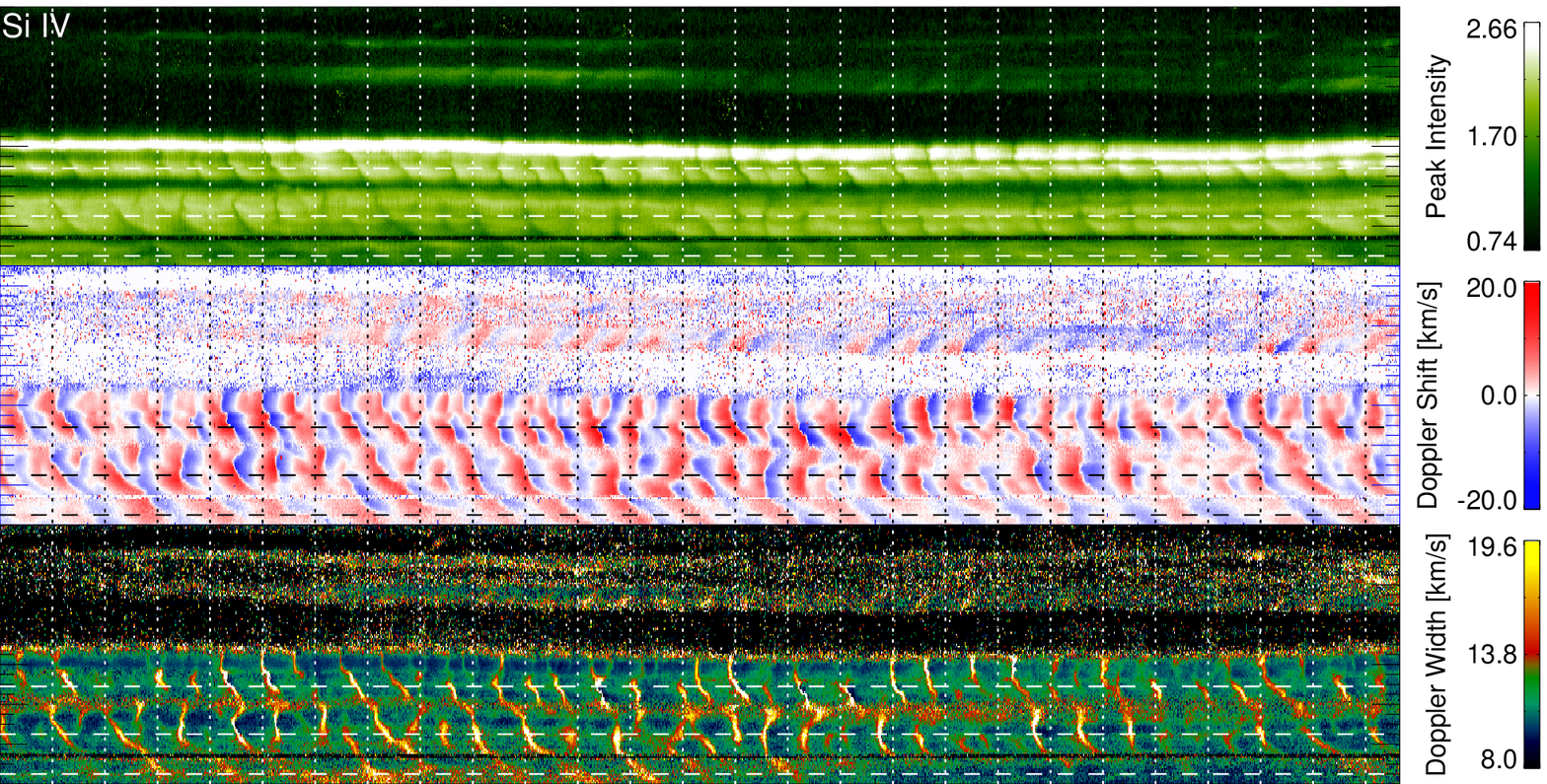}} 
\end{minipage}
\begin{minipage}[t]{0.6\textheight}
{\includegraphics[width=\textwidth]{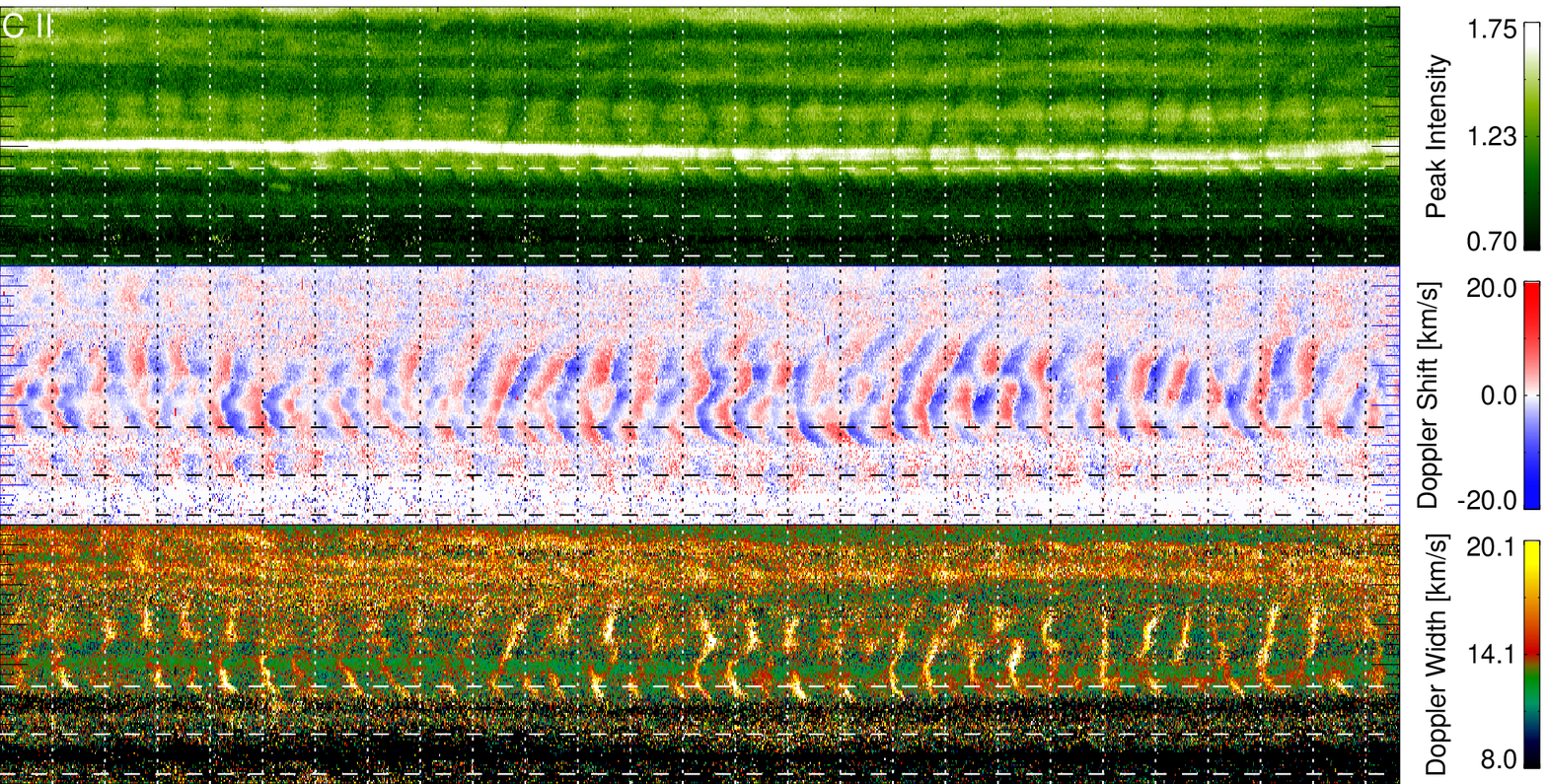}} 
\end{minipage}
\begin{minipage}[t]{0.6\textheight}
{\includegraphics[width=\textwidth]{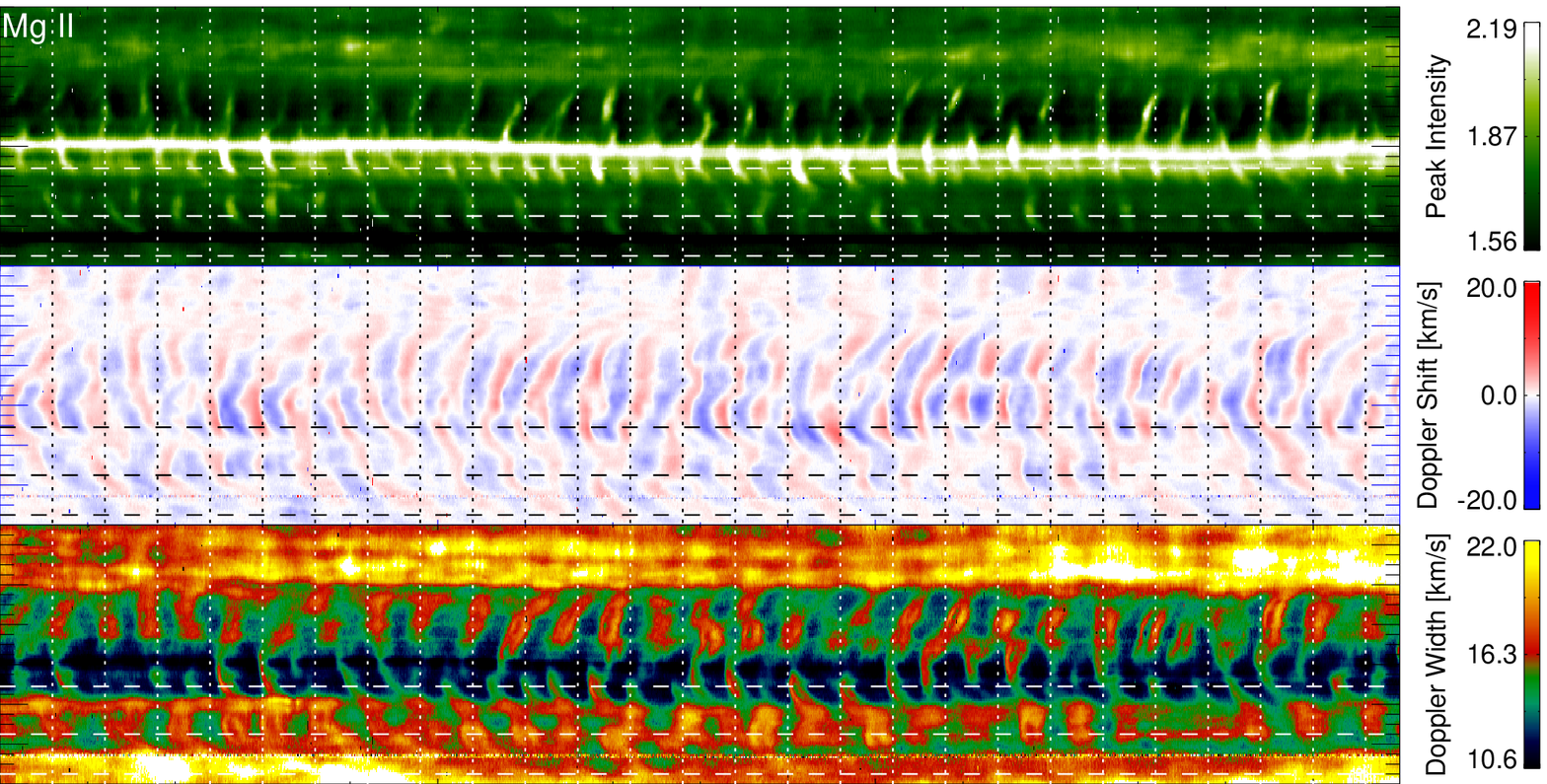}} 
\end{minipage}
\caption{ Temporal evolution of the SGF parameters in the spatial range indicated by the two long horizontal lines in Figure~\ref{fig.1}. From top to bottom: Si~{\sc{iv}}, C~{\sc{ii}} and Mg~{\sc{ii}}. The length along the slit (vertical dimension) is 22$^{\prime\prime}$. The total duration (horizontal dimension) is 80 minutes. The interval between two vertical dotted lines is 3 minutes. The three horizontal lines in each panel indicate the slit locations 1, 2, \& 3 in Figure~\ref{fig.1}. Note that data points with low signal to noise ratio are shown in white and black in the images of Doppler shift and line width (for Si~{\sc{iv}} and C~{\sc{ii}}), respectively.}
\label{fig.2}
\end{figure*}

The normally central-reversed Mg~{\sc{ii}}~and C~{\sc{ii}}~line profiles have almost no central reversal in the sunspot. A similar phenomenon has also been reported for the Hydrogen Lyman lines and it suggests a greatly reduced opacity in the sunspot atmosphere \citep{Tian2009}. Since profiles of all the three lines within the sunspot are mostly not reversed and close to Gaussian to some extent, as a first step we applied a single Gaussian fit (SGF) to all line profiles in the slit range between the two long horizontal lines in Figure~\ref{fig.1}(D)-(I). As a reference, each line was assumed to have a zero Doppler shift on average. The temporal evolution of the SGF parameters are shown in Figure~\ref{fig.2}. 

From Figure~\ref{fig.2} we can see the oscillation pattern in not only the intensity, but also the Doppler shift and width of each line. Inclined stripe-like structures, indicative of phase difference at different slit locations, are present on these maps. The oscillation at the center of the selected slit portion clearly precedes the upper and lower parts of the slit. The SJI 1400\AA{} movie shows an apparent outward propagation for the waves. Using the slopes of the inclined stripe-like structures, the speed of the apparent motion is estimated to be around 25~km~s$^{-1}$. This value is comparable to the sound speed and much smaller than the Alfv\'en speed \citep{OShea2002,McIntosh2011} in the TR. 

\begin{figure*}
\centering {\includegraphics[width=\textwidth]{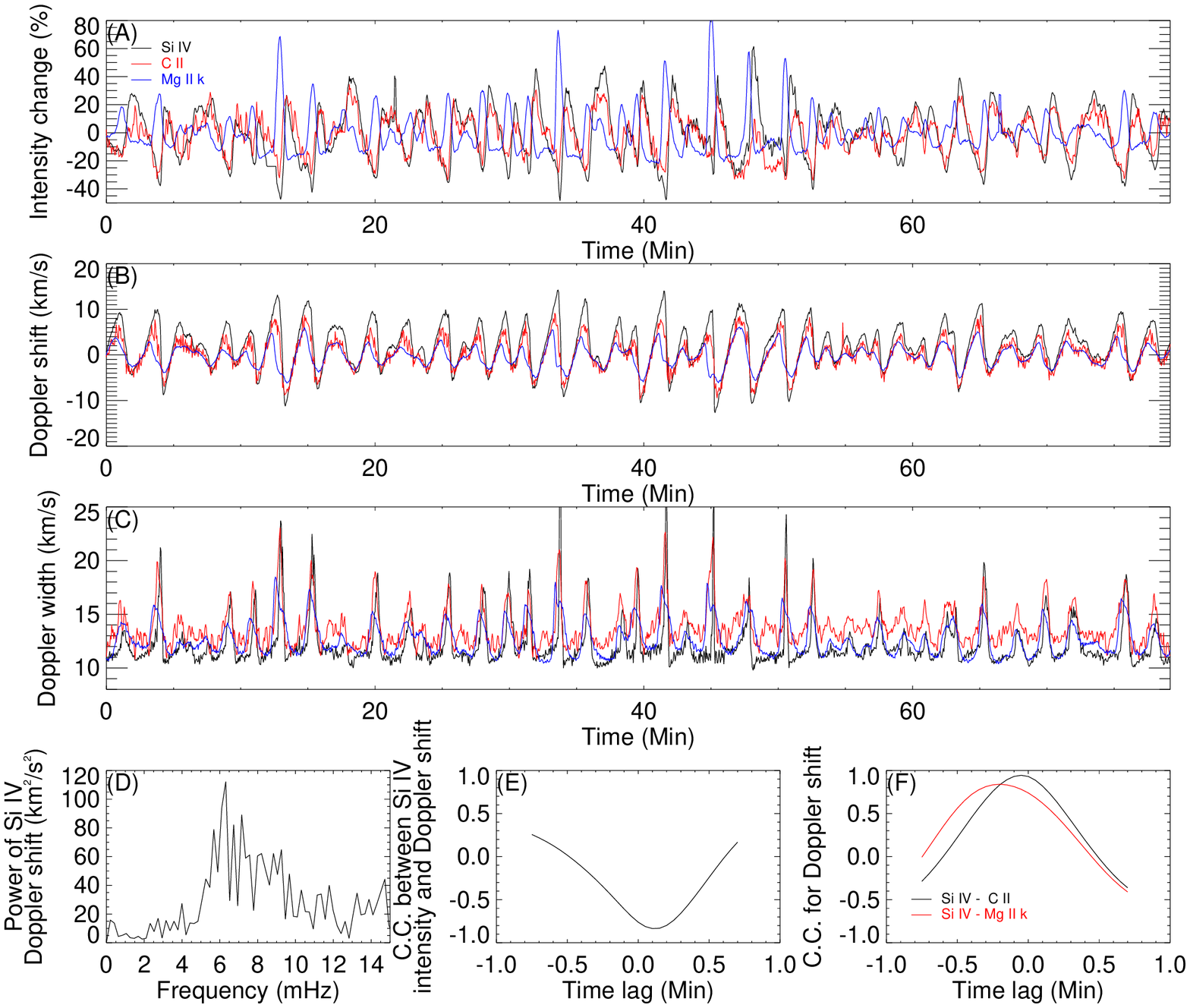}} \caption{ (A)-(C) Temporal evolution of the SGF peak intensity, Doppler shift, and line width at the slit location 3. (D) Fourier power spectrum of the Si~{\sc{iv}} Doppler shift. (E) Correlation coefficient between the intensity and Doppler shift of Si~{\sc{iv}} as a function of time lag. (F) Correlation coefficient between the Si~{\sc{iv}} Doppler shift and the Doppler shift of Mg~{\sc{ii}}/C~{\sc{ii}} as a function of time lag. }
\label{fig.3}
\end{figure*}

The oscillation is largely nonlinear. Evidence includes the large intensity change and the sharp change from red shift to blue shift. As an example, Figure~\ref{fig.3}(A)-(C) shows the temporal evolution of the SGF parameters at the slit location 3. The intensity is shown as percentage after constructing a background through a 5-minute boxcar smoothing of the time series of intensity at each spatial pixel. The intensity change is of the order of 30\%. The Doppler shift oscillation reveals a sawtooth pattern with an amplitude of about 10~km~s$^{-1}$ in Si~{\sc{iv}}. These values are at least three times larger than those measured from the C~{\sc{iv}}~1548.19\AA{} line in several sunspot oscillations \citep{Gurman1982}. The velocity amplitude is slightly smaller in C~{\sc{ii}} and Mg~{\sc{ii}}. Correlated changes between intensity and Doppler shift are often interpreted as propagating slow mode magnetoacoustic waves \citep[e.g.,][]{Maltby1999,Sakurai2002,Wang2009,Nishizuka2011}. Figure~\ref{fig.3} reveals a correlation between blue shift and enhanced intensity of the TR lines, suggesting the possible presence of upward propagating magneto-acoustic shock waves. Figure~\ref{fig.3}(E) also shows that the intensity oscillation of Si~{\sc{iv}} slightly lags the Doppler shift oscillation, suggesting that the intensity increase is caused by the plasma compression after the impulsive ascending motion of the shocks. 

The Doppler shift oscillation of Si~{\sc{iv}} lags those of C~{\sc{ii}} and Mg~{\sc{ii}} by $\sim$3 and $\sim$12 seconds, respectively (Figure~\ref{fig.3}(F)). Considering the different formation heights of the three lines, these time lags also support the interpretation of upward propagating waves from the chromosphere to the TR. We noticed that \cite{Brynildsen1999b} reached a similar conclusion based on a time lag of $\sim$25 seconds for the sunspot oscillation in N~{\sc{v}} relative to that in Si~{\sc{ii}}.  \cite{OShea2002} also found time lags of 16-68 seconds for sunspot oscillations in TR lines relative to the oscillations in UV continuum formed at temperature minimum. These measured time lags will provide constraint to numerical simulations of sunspot oscillations. Our measured time lags also suggest that the formation heights of Si~{\sc{iv}}  and C~{\sc{ii}} are very close, whereas Mg~{\sc{ii}} forms much lower in the sunspot atmosphere. 

The periods of the oscillations are mainly in the range of 95-180 s at slit location 3 (Figure~\ref{fig.3}(D)) and slightly larger at location 2. The periods are in the range of 200-320s at slit location 1. The increase of the oscillation period with distance from the sunspot center is consistent with the scenario of more inclined magnetic filed lines towards the outer part of the sunspot: the acoustic cutoff frequency is reduced and thus waves with longer periods can propagate along these inclined magnetic field lines into the chromosphere and TR \citep[]{Jess2013,Yuan2014}. 

\begin{figure*}
\centering {\includegraphics[width=\textwidth]{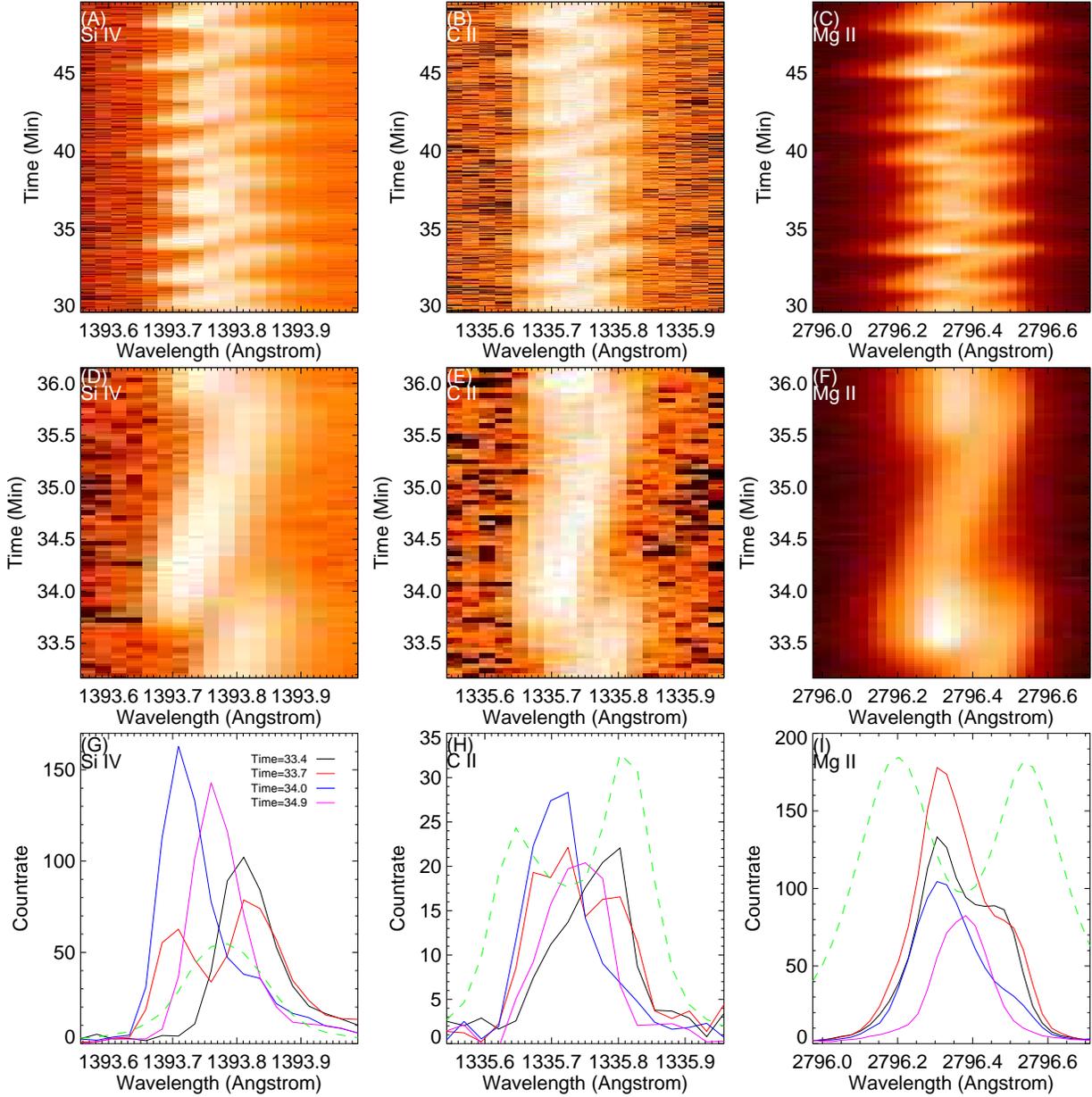}} \caption{ (A)-(C) Wavelength-time plots for Si~{\sc{iv}}, C~{\sc{ii}} and Mg~{\sc{ii}} at the slit location 3 in part of the time range. (D)-(F) Same as (A)-(C) but for a shorter time range. (G)-(I): Profiles of the three lines at several times. Typical profiles in the plage are shown as the dashed lines. }
\label{fig.4}
\end{figure*}

Figure~\ref{fig.3}(C) reveals an in-phase line width oscillation for the three lines. The line width of Si~{\sc{iv}} is almost constant most of the time and suddenly increases by a factor of two when the Doppler shift changes sharply from red to blue. Similar line width oscillations in sunspots have also been reported by \cite{Brynildsen1999b}. However, we found that this nonlinear line width oscillation is related to periodic occurrence of multiple components after checking the line profiles. Figure~\ref{fig.4}(G)-(I) clearly show that the line profiles exhibit two distinct components during the impulsive change of Doppler shift. These two components likely represent the emission from the newly shocked plasma and the back-falling material after the passage of the previous shock, respectively. A SGF to such line profiles would certainly lead to a smaller intensity and largely enhanced line width as evident in Figures~\ref{fig.2} and \ref{fig.3}. Obviously, such an increase in the line width is not caused by enhanced heating or turbulence. Similar effects have also been reported in coronal outflows and coronal mass ejections \citep[e.g.,][]{DePontieu2009,McIntosh2009a,McIntosh2009b,Bryans2010,Tian2011,Tian2012}. One needs to be cautious when interpreting the line parameters derived from SGF. We noticed that \cite{Centeno2006} observed irregular shapes of the Stokes V profiles of chromospheric lines during the red- to blue-shift transition. They also suggested the integration along the line of sight in the shocked plasma as a possible explanation. Our high-resolution IRIS observations thus likely reveal an interesting behavior of the shocks associated with sunspot oscillations - a new shock occurs before the complete fading of the previous shock. The observed line emission is thus from two different sources, one with downflowing plasma and the other with upflowing plasma. Such a behavior should be taken into account in models of sunspot oscillations.

The temporal evolution of the line profiles at slit location 3 is plotted as the wavelength-time maps in Figure~\ref{fig.4}(A)-(F). The motion of the line core is dominated by the following behavior: a fast blueward excursion accompanied by an intensity increase, followed by a gradual redward excursion. Such behavior has been reported for the chromospheric Ca~{\sc{ii}} H\&K and He~{\sc{i}}~10830\AA{} lines in sunspots \citep[e.g.,][]{Rouppe2003,Centeno2006,Felipe2010} and was attributed to shock waves by \cite{Lites1986}. We noticed that \cite{Brynildsen1999b} reported similar behavior for several TR lines, which were attributed to the presence of nonlinearities and possibly shocks. However, the low resolution and low cadence of the data they analyzed did not allow them to reach a solid conclusion on the nature of the waves.  Recently, \cite{Bard2010} performed radiation hydrodynamic simulations of the formation and evolution of the Ca~{\sc{ii}} H\&K lines in a sunspot. They claimed that umbral flashes result from increased emission of the local material during the passage of acoustic waves originating in the photosphere and steepening to shock in the chromosphere. 

Our IRIS observations reveal some clear differences between the chromopheric and TR lines, although the general shock behavior described above is similar for all the three lines. Figure~\ref{fig.4}(D)-(F) shows that the shock signature appears first in the Mg~{\sc{ii}} line and then in the TR lines. This time lag is consistent with Figure~\ref{fig.3}(F). For the Mg~{\sc{ii}} line, the intensity enhancement occurs before the maximum blue shift is reached. Similar behavior was also found for the Ca~{\sc{ii}} H\&K lines by \cite{Rouppe2003}. However, for the TR lines the maximum intensity occurs slightly later than the maximum blue shift. The different behaviors might be explained by the fact that Mg~{\sc{ii}} is optically much thicker than the TR lines. In that case, the intensity is determined by not only the density but also the temperature. These different behaviors may provide important constraint to models of sunspot oscillations. 

\begin{figure*}
\centering {\includegraphics[width=\textwidth]{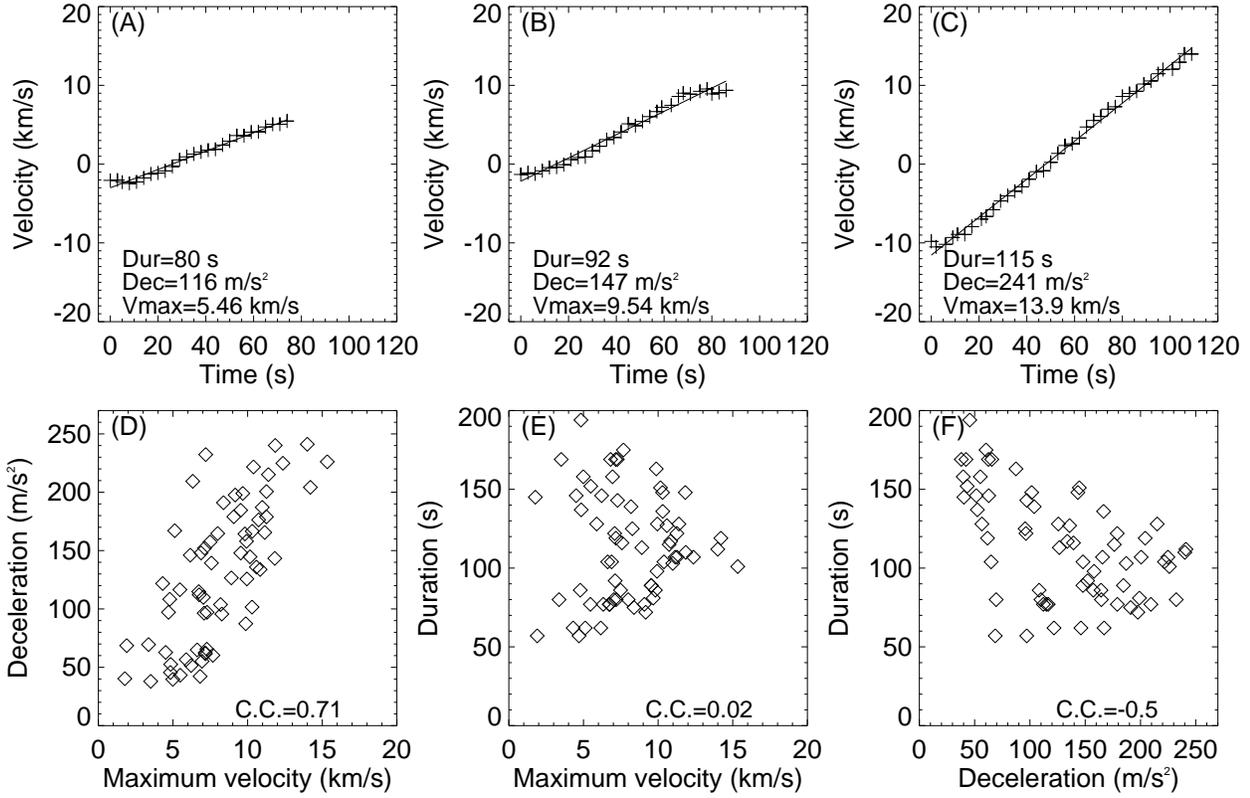}} \caption{ (A)-(C) Examples of linear fit to the velocity-time relationship in the deceleration phase of shocks. The duration (Dur), maximum velocity (Vmax) and deceleration (Dec) are shown in each panel. (D)-(F) Scatter plots of the relationship between shock parameters. The correlation coefficients are also shown in each panel. }
\label{fig.5}
\end{figure*}

The behavior shown in Figure~\ref{fig.4}(C) has also been reported in the Ca~{\sc{ii}} and H$\alpha$ lines for dynamic fibrils (DFs) in plages \citep[e.g.,][]{Langangen2008a,Langangen2008b}. Through radiative MHD simulations, \cite{Hansteen2006} and \cite{DePontieu2007} found that the highly dynamic chromospheric shock waves cause significant up- and downward excursions of the upper chromosphere. The velocity profile shows an impulsive acceleration at the beginning of the ascending phase and constant deceleration afterwards. The striking similarities between observation and simulation results led them to conclude that DFs are driven by magneto-acoustic shocks. The simulations predict that DFs with a larger deceleration show a larger maximum velocity, which has also been observed.

Although these simulations deal with DFs, shock behavior should be the same even when we look at sunspots. We thus identified 67 shocks from the temporal evolution of the Doppler shift (e.g., Figure~\ref{fig.3}(B)) of the optically thin Si~{\sc{iv}}~line at slit locations 1-3, and performed a linear fit to the velocity profile in the constant deceleration phase of each shock. Here we define the constant deceleration phase of one shock as the time interval after the impulsive acceleration, thus including both the decelerating ascending motion and the subsequent receding motion. The constant deceleration was calculated from the fitting. Figure~\ref{fig.5} shows three examples of the fitting results and the scatter plots of the relationship between shock parameters (duration, maximum velocity and deceleration of the deceleration phase). Note that only a few relatively obvious shocks were identified at location 1. These longer-period (200-320s) oscillations are probably related to the RPWs and the shock signatures described above seem to be much weaker compared to those at locations 3\&2, where the oscillations have a period around 3 minutes.  

There is a clear correlation between the deceleration and maximum velocity, with a linear Pearson correlation coefficient of 0.71. This correlation is very similar to what has been found in DFs and provides additional evidence that the sunspot oscillations are upward propagating shock waves. The presence of shock signatures in the Si~{\sc{iv}}~line demonstrates that the shock waves formed in the chromosphere can penetrate into the TR. We noticed that the maximum velocities are comparable to those of DFs in Doppler measurements by \cite{Langangen2008a,Langangen2008b}, but 2-3 times smaller than those of DFs and  and quiet-Sun mottles measured from the proper motions in imaging observations \citep{Hansteen2006,DePontieu2007,Rouppe2007}. The average deceleration, however, is much larger than the value of DFs in Doppler measurements and comparable to that measured from the proper motions. These differences are partly related to the different viewing angles in different observations. The observed values of the deceleration are only a fraction of solar gravity and incompatible with a ballistic path at solar gravity. In contrast, these values fit very well with the shock wave deceleration model of \cite{Hansteen2006}.

The correlation between the duration and deceleration/maximum velocity, also predicted by \cite{DePontieu2007}, is weaker or not obvious. Similar results have also been reported by \cite{Langangen2008a} in DFs. One possible explanation is the projection effect: the measured duration does not suffer from projection effects, whearas the deceleration and velocity do. On the other hand, these correlations are weak in some simulations \citep{Heggland2007,Martinez-Sykora2009}. Nevertheless, from Figure~\ref{fig.5}(F) we see a general trend of smaller acceleration for longer duration. Again this relationship is incompatible with a ballistic model but fits very well with the shock wave deceleration model \cite{DePontieu2007}. 

The Wentzel-Kramers-Brillouin (WKB) energy flux of slow magneto-acoutsic waves can be estimated in the following way \citep[e.g.,][]{Ofman1999}:

\begin{equation}
\emph{$F=0.5\rho(\delta v)^{2}C_s$}\label{equation1},
\end{equation}

\noindent where $\rho$ is the mass density. Using the density-sensitive line pair O~{\sc{iv}}~1399.77\AA{}/1401.16\AA{},  we obtained an electron density of log ({\it N}$_{e}$/cm$^{-3}$)=10.75 using CHIANTI v7.0 \citep{Dere1997,Landi2012}. Taking the observed value of the velocity amplitude $\delta v$=10 km~s$^{-1}$ and the sound speed of 38 km~s$^{-1}$ calculated using the formation temperature of Si~{\sc{iv}}, we obtain an energy flux of 1.8$\times$10$^5$ erg cm$^{-2}$ s$^{-1}$, which is one order of magnitude lower than that required to balance the radiative and conductive losses of the active corona (2$\times$10$^6$ erg cm$^{-2}$ s$^{-1}$). The energy flux will increase if we take into account the projection effects. We noticed that the energy flux we obtained here is only 40\% smaller than that estimated from chromospheric lines in MHD simulations \citep{Felipe2011}, which suggests that a significant amount of the energy that reaches the high chromosphere is transmitted to the TR. But we have to remember that the density might be overestimated by up to a factor of 10 due to the effect of non-equilibrium ionization \citep{Olluri2013}.

We have also analyzed another sit-and-stare observation of sunspot made with IRIS from 17:11 to 17:35 on 2013 July 20. The pointing of this observation is (46$^{\prime\prime}$, 288$^{\prime\prime}$). The different timing of the maximum intensity and blue shift for TR and chromospheric lines, the spurious line width oscillation, and the strong correlation between deceleration and maximum velocity are also clearly revealed in this observation. Results for this observation are not shown here since they are basically the same as those shown above.

\section{Conclusion}
We report the first results of sunspot oscillations observed with IRIS. Our results reveal several new aspects of the shock wave behavior for sunspot oscillations in the TR and chromosphere.

We first apply a single Gaussian fit to the profiles of the Mg~{\sc{ii}}~2796.35\AA{}, C~{\sc{ii}}~1335.71\AA{} and Si~{\sc{iv}}~1393.76\AA{} lines in the sunspot since these line profiles are mostly close to Gaussian to some extent. The intensity oscillation has an amplitude of $\sim$30\% for all three lines. The Doppler shift oscillation reveals a sawtooth pattern with an amplitude of about 10~km~s$^{-1}$ in Si~{\sc{iv}} and slightly smaller values in C~{\sc{ii}}~and Mg~{\sc{ii}}. The Si~{\sc{iv}} oscillation lags the C~{\sc{ii}} and Mg~{\sc{ii}} oscillations by about 3 and 12 seconds, respectively. The correlated change between the intensity and blue shift of the TR lines, together with the time lags and the strong nonlinearities, suggest the presence of magneto-acoustic shock waves propagating from the chromosphere to the TR.

Detailed analysis of the temporal evolution of the line profiles reveals a repeated pattern of the following behavior: the line core first experiences a sudden impulsive blueward excursion and an accompanied intensity enhancement, then starts a gradual and constant deceleration to the red side. The maximum red shift is correlated with the deceleration. Such behaviors have been previously found in the chromospheric emission of DFs and proven to be the signatures of upward propagating magneto-acoustic shock waves. The similar behavior in both the chromospheric and TR emission lines in our data suggests that the three-minute sunspot oscillations may be dominated by a similar process. Waves generated by convective flows and global p-mode oscillations in the photosphere leak upward, steepen, and form shocks in the sunspot chromosphere and TR. A plasma parcel passing through a shock will then experience a sudden impulse ascending motion, followed by a gradual and constant deceleration. We also found that the maximum intensity slightly lags the maximum blue shift for the TR lines. However, the intensity enhancement of Mg~{\sc{ii}} occurs before the maximum blue shift is reached.  

We have also demonstrated that the strongly nonlinear line width oscillation, observed both here and previously, is actually related to the superposition of multiple emission components. The line profiles clearly exhibit two distinct components during the impulsive change from red shift to blue shift. These two components represent the emission from the newly shocked plasma and the back-falling matter after the passage of the previous shock, respectively. They are likely caused by the behavior of the shocks - a new shock occurs before the complete fading of the previous shock. The greatly enhanced SGF line width is mainly caused by the superposition of the two emission components.

\begin{acknowledgements}
IRIS is a NASA small explorer mission developed and operated by LMSAL with mission operations executed at NASA Ames Research center and major contributions to downlink communications funded by the Norwegian Space Center (NSC, Norway) through an ESA PRODEX contract. This work is supported by NASA under contract NNG09FA40C (IRIS) and the Lockheed Martin Independent Research Program, the European Research Council grant agreement No. 291058, and contract 8100002705 from LMSAL to SAO. H. T. thanks Luc Rouppe van der Voort and Jorrit Leenaarts for useful discussion. 
\end{acknowledgements}


\begin{thebibliography}{}
%
\bibitem[Bard \& Carlsson(1997)]{Bard1997}
Bard, S., Carlsson, M. 1997, ESA Special Publication, 404, 189
%
\bibitem[Bard \& Carlsson(2010)]{Bard2010}
Bard, S., Carlsson, M. 2010, ApJ, 722, 888
%
\bibitem[Beckers \& Tallant(1969)]{Beckers1969}
Beckers, J. M., \& Tallant, P. E. 1969, Sol. Phys., 7, 351
%
\bibitem[Beckers \& Schultz(1972)]{Beckers1972}
Beckers, J. M., \& Schultz, R. B. 1972, Sol. Phys. 27, 61
%
\bibitem[Bloomfield et al.(2007)]{Bloomfield2007}
Bloomfield, D. S., Lagg, A., \& Solanki, S. K. 2007, ApJ, 671, 1005
%
\bibitem[Bogdan(2000)]{Bogdan2000}
Bogdan, T. J. 2000, Sol. Phys., 192, 373
%
\bibitem[Bogdan \& Judge(2006)]{Bogdan2006}
Bogdan, T. J., \& Judge, P. G. 2006, Philos. Trans. R. Soc. London A, 364, 313
%
\bibitem[Brekke(1993)]{Brekke1993}
Brekke, P. 1993, ApJS, 87, 443
%
\bibitem[Brynildsen et al.(1999a)]{Brynildsen1999a}
Brynildsen, N., Leifsen, T., Kjeldseth-Moe, O., Maltby, P., 1999a, ApJ, 511, L121
%
\bibitem[Brynildsen et al.(1999b)]{Brynildsen1999b}
Brynildsen, N., Kjeldseth-Moe, O., Maltby, P., \& Wilhelm, K. 1999b, ApJ, 517, L159
%
\bibitem[Brynildsen et al.(2001)]{Brynildsen2001}
Brynildsen, N., Maltby, P., Kjeldseth-Moe, O., Wilhelm, K. 2001, A\&A, 552, L77
%
\bibitem[Brynildsen et al.(2002)]{Brynildsen2002}
Brynildsen, N., Maltby, P., Fredvik, T., \& Kjeldseth-Moe, O. 2002, Sol. Phys., 207, 259
%
\bibitem[Brynildsen et al.(2004)]{Brynildsen2004}
Brynildsen, N., et al. 2004, Sol. Phys., 221, 237
%
\bibitem[Bryans et al.(2010)]{Bryans2010}
Bryans, P., Young, P. R., \& Doschek, G. A. 2010, ApJ, 715, 1012
%
%
\bibitem[Centeno et al.(2006)]{Centeno2006}
Centeno, R., Collados, M., Trujillo Bueno, J. 2006, ApJ, 640, 1153
%
%
\bibitem[Dere et al.(1997)]{Dere1997}
Dere, K. P., Landi, E., Mason, H. E., Monsignori-Fossi, B. C., \& Young, P. R. 1997, A\&AS, 125, 149
%
\bibitem[de la Cruz Rodr\'{\i}guez et al.(2013)]{delaCruz2013}
de la Cruz Rodr\'{\i}guez, J., Rouppe van der Voort, L., Socas-Navarro, H., van Noort, M. 2013, A\&A, 556, A115
%
\bibitem[De Moortel et al.(2002)]{DeMoortel2002}
De Moortel, I., Ireland, J., Hood, A. W., \& Walsh, R. W. 2002, A\&A, 387, L13
%
\bibitem[De Pontieu et al.(2004)]{DePontieu2004}
De Pontieu, B., Erd\'{e}lyi, R., \& James, S. P. 2004, Nature, 430, 536
%
%
\bibitem[De Pontieu et al.(2007)]{DePontieu2007}
De Pontieu, B., Hansteen, V. H., Rouppe van der Voort, L., van Noort, M., Carlsson, M. 2007, ApJ, 655, 624
%
\bibitem[De Pontieu et al.(2009)]{DePontieu2009}
De Pontieu, B., McIntosh, S. W., Hansteen, V. H., \& Schrijver, C. J., 2009, ApJ, 701, L1
%
\bibitem[De Pontieu et al.(2014)]{DePontieu2014}
De Pontieu, B., et al. 2014, Sol. Phys., 289, 2733
%
\bibitem[Felipe et al.(2010)]{Felipe2010}
Felipe, T., Khomenko, E., Collados, M., Beck, C. 2010, ApJ, 722, 131
%
\bibitem[Felipe et al.(2011)]{Felipe2011}
Felipe, T., Khomenko, E., Collados, M. 2011, ApJ, 735, 65
%
\bibitem[Foukal et al.(1974)]{Foukal1974}
Foukal, P. V., Huber, M. C. E., Noyes, R. W., et al. 1974, ApJ, 193, L143
%
\bibitem[Giovanelli et al.(1972)]{Giovanelli1972}
Giovanelli, R. G. 1972, Sol. Phys., 27, 71
%
\bibitem[Gurman et al.(1982)]{Gurman1982}
Gurman, J. B., Leibacher, J. W., Shine, R. A., Woodgate, B. E., Henze, W. 1982, ApJ, 253, 939
%
\bibitem[Hansteen et al.(2006)]{Hansteen2006}
Hansteen, V. H., De Pontieu, B., Rouppe van der Voort, L., van Noort, M., Carlsson, M. 2006, ApJ, 647, L73
%
\bibitem[Heggland et al.(2007)]{Heggland2007}
Heggland, L., De Pontieu, B., Hansteen, V. H. 2007, ApJ, 666, 1277
%
\bibitem[Jess et al.(2013)]{Jess2013}
Jess, D. B., Reznikova, V. E., Van Doorsselaere, T., Keys, P. H., Mackay, D. H. 2013, ApJ, 779, 168
%
\bibitem[Landi et al.(2012)]{Landi2012}
Landi, E., Del Zanna, G., Young, P. R., Dere, K. P., Mason, H. E. 2012, ApJ, 744, 99
%
\bibitem[Langangen et al.(2008a)]{Langangen2008a}
Langangen, \O., Carlsson, M., Rouppe van der Voort, L., Hansteen, V., De Pontieu, B. 2008a, ApJ, 673, 1194
%
\bibitem[Langangen et al.(2008b)]{Langangen2008b}
Langangen, \O., Rouppe van der Voort, L., Lin, Y. 2008b, ApJ, 673, 1201
%
\bibitem[Lemen et al.(2012)]{Lemen2012}
Lemen, J. R., et al. 2012, Solar Phys., 275, 17
%
\bibitem[Lites(1986)]{Lites1986}
Lites, B. W. 1986, ApJ, 301, 1005
%
\bibitem[Lites(1992)]{Lites1992}
Lites, B. W. 1992, in Sunspots: Theory and Observations, ed. J. H. Thomas \& N. O. Weiss (NATO ASI Ser. C, 375; Dordrecht: Kluwer), 261
%
\bibitem[Mart\'{\i}nez-Sykora et al.(2009)]{Martinez-Sykora2009}
Mart\'{\i}nez-Sykora, J., Hansteen, V., De Pontieu, B., Carlsson, M. 2009, ApJ, 701, 1569
%
\bibitem[Maltby et al.(1999)]{Maltby1999}
Maltby, P, et al. 1999, Solar Phys., 190, 437
%
\bibitem[McIntosh \& De Pontieu(2009a)]{McIntosh2009a}
McIntosh, S. W., \& De Pontieu, B. 2009a, ApJ, 706, L80 
%
\bibitem[McIntosh \& De Pontieu(2009b)]{McIntosh2009b}
McIntosh, S. W., \& De Pontieu, B. 2009b, ApJ, 707, 524
%
\bibitem[McIntosh et al.(2011)]{McIntosh2011}
McIntosh, S. W., De Pontieu, B., Carlsson, M., et al. 2011, Nature, 475, 477
%
%
\bibitem[Nishizuka \& Hara(2011)]{Nishizuka2011}
Nishizuka, N., \& Hara, H. 2011, ApJ, 737, L43
%
\bibitem[Ofman et al.(1999)]{Ofman1999}
Ofman, L., Nakariakov, V. M., \& DeForest, C. E. 1999, ApJ, 514, 441
%
\bibitem[Olluri et al.(2013)]{Olluri2013}
Olluri, K., Gudiksen, B. V., Hansteen, V. H. 2013, ApJ, 767, 43
%
\bibitem[O$^{\prime}$Shea et al.(2002)]{OShea2002}
O$^{\prime}$Shea, E., Muglach, K., \& Fleck, B. 2002, A\&A, 387, 642
%
\bibitem[Pesnell et al.(2012)]{Pesnell2012}
Pesnell, W. D., Thompson, B. J., Chamberlin, P. C. 2012, Sol. Phys., 275, 3
%
\bibitem[Rouppe van der Voort et al.(2003)]{Rouppe2003}
Rouppe van der Voort, L. H. M., Rutten, R. J., S\"utterlin, P., Sloover, P. J., Krijger, J. M. 2003, A\&A, 403, 277
%
\bibitem[Rouppe van der Voort et al.(2007)]{Rouppe2007}
Rouppe van der Voort, L. H. M., De Pontieu, B., Hansteen, V. H., Carlsson, M., van Noort, M. 2007, ApJ, 660, L169
\bibitem[Sakurai et al.(2002)]{Sakurai2002}
Sakurai, T., Ichimoto, K., Raju, K. P., Singh, J. 2002, Sol. Phys., 209, 265
%
\bibitem[Thomas(1985)]{Thomas1985}
Thomas, J. H. 1985, Australian Journal of Physics, 38, 811
%
\bibitem[Thomas et al.(1987)]{Thomas1987}
Thomas, J. H., Lites, B. W., Gurman, J. B., Ladd, E. F. 1987, ApJ, 312, 457
%
%
\bibitem[Tian et al.(2009)]{Tian2009}
Tian, H., Curdt, W., Teriaca, L., Landi, E., \& Marsch, E. 2009, A\&A, 505, 307
%
\bibitem[Tian et al.(2011)]{Tian2011}
Tian, H., McIntosh, S. W., De Pontieu, B., Mart\'{\i}nez-Sykora, J., Sechler, M., \& Wang, X. 2011, ApJ, 738, 18
%
\bibitem[Tian et al.(2012)]{Tian2012}
Tian, H., McIntosh, S. W., Xia, L.-D., He, J.-S., \& Wang, X. 2012, ApJ, 748, 106
%
%
\bibitem[Tziotziou et al.(2006)]{Tziotziou2006}
Tziotziou, K., Tsiropoula, G., Mein, N., \& Mein, P. 2006, A\&A, 456, 689
%
%
\bibitem[Wang et al.(2009)]{Wang2009}
Wang, T.-J., Ofman, L., \& Davila, J. M. 2009, ApJ, 696, 1448
%
\bibitem[Yuan et al.(2014)]{Yuan2014}
Yuan, D., Sych, R., Reznikova, V. E., Nakariakov, V. M. 2014, A\&A, 561, A19
%
%
\bibitem[Zirin \& Stein(1972)]{Zirin1972}
Zirin, H., \& Stein, A. 1972, ApJL, 178, L85
%
\end{thebibliography}
\end{document}